\documentclass{ws-mpla}
\usepackage[super]{cite}
\usepackage{graphicx}

\usepackage{amssymb}
\usepackage{amsmath}
\usepackage{float}
\usepackage{trfsigns}

\usepackage{tensor}
\newcommand{\te}[2]{\tensor{#1}{#2}}

\newcommand \be{\begin{eqnarray}}
\newcommand \ee{\end{eqnarray}}

\newcommand \ba{\begin{align}}
\newcommand \eea{\end{align}}

\begin{document}

\markboth{Morawetz}{Time behaviour of Hubble parameter by torsion}

\catchline{}{}{}{}{}

\title{Time behaviour of Hubble parameter by torsion
}

\author{Klaus Morawetz
}
\address{M\"unster University of Applied Sciences,
Stegerwaldstrasse 39, 48565 Steinfurt, Germany\\
International Institute of Physics- UFRN,
Campus Universit\'ario Lagoa nova,
59078-970 Natal, Brazil
\\
morawetz@fh-muenster.de}



\maketitle

\pub{Received (Day Month Year)}{Revised (Day Month Year)}

\begin{abstract}
Consequences of the consistent exact solution of Einstein-Cartan equation on the time dependence of Hubble parameter are discussed. The torsion leads to a space and time dependent expansion parameter which results into nontrivial windows of Hubble parameter between diverging behaviour. Only one window shows a period of decreasing followed by increasing time dependence. Provided a known cosmological constant and the present values of Hubble and deceleration parameter this changing time can be given in the past as well as the ending time of the windows or universe. The comparison with the present experimental data allows to determine all parameters of the model. Large-scale spatial periodic structures appear. From the metric with torsion outside matter it is seen that torsion can feign dark matter. 
\end{abstract}

\ccode{PACS No: 04.50.Kd, 
98.80.Es, 
98.80.-k 
}

\section{Introduction}

There is an ongoing discrepancy of Hubble data from the early universe obtained by background radiation and data from present galaxies \cite{Riess22}:
"We find a 5$\sigma$ difference with the prediction of $H_0$ from Planck cosmic microwave background
observations under $\lambda$CDM, with no indication that the discrepancy arises from measurement uncertainties or
analysis variations considered to date. The source of this now long-standing discrepancy between direct and
cosmological routes to determining $H_0$ remains unknown." This discrepancy is also further supported by quasars at far distance \cite{RL19,Li21}. This tension in $H_0$ is related to the anomaly in $\Omega_{m}$ of the Friedmann equations. $H_0$ decreases with effective redshift, while $\Omega_{m}$ increases with effective redshift in LCDM models \cite{CJSBCDS22}. This is attributed to negative dark energy density \cite{MCCPS23}. Obviously only in special models $H_0$ does not evolve.

\begin{figure}
\centerline{\includegraphics[width=8.5cm]{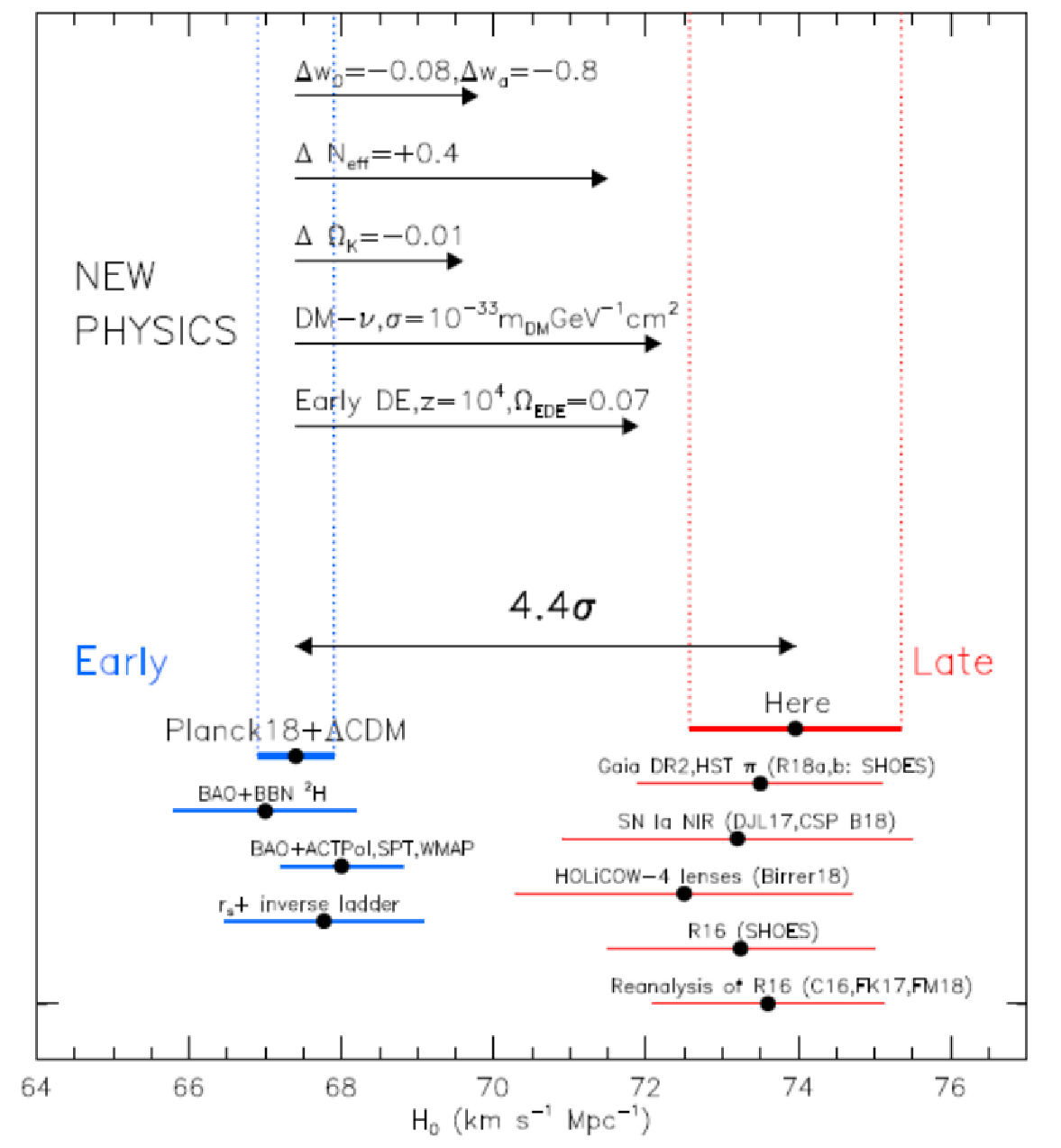}}
\caption{\label{hubble_riess} The present data of Hubble constant from \cite{Riess19}.}  
\end{figure}

In figure~\ref{hubble_riess} the data of the earlier paper are presented and obviously the discrepancy has increased now by 5$\sigma$. 

Considering the time dependence of the scale parameter
\be
R(t)=R_0(t_0)[1+H_0(t-t_0)-\frac q 2 H^2(t-t_0)^2+...]
\ee
with the Hubble and deceleration parameters 
\be
H&=&{\dot R \over R}\qquad q=-{\ddot R R\over R^2}
\label{hq}
\ee
we adopt for the moment (\ref{hq}) as a literal definition which would lead to a time change of Hubble parameter
\be
\dot H={\ddot R R-{\dot R}^2\over R^2}.
\ee
Demanding $\dot H >0$ means $\ddot R R>{\dot R}^2$ and with the help of (\ref{hq}) one finds the equivalence
\be
\dot H >0 \,\leftrightarrow \, q<-1.
\ee
Consequently the search is going on for such a negative deceleration parameter. A analysis of the Planck data and SHOES collaboration \cite{CM20} indeed
seems to indicate parameters
\be
H_0&=&75.35\pm 1.68 {\rm km/sMpc}\nonumber\\
q_0&=&-1.08\pm 0.29.
\label{values}
\ee

The theoretical challenge is how the Hubble parameter could increase with time \cite{CMY19,MZJ20,CJ21}. One possible solution is provided by torsion leading to a metric due to an exact solution \cite{Mo21,Mo21e} of Einstein-Cartan equations. In the latter the additional gravitational potential due to torsion becomes
\be
U_{ik}={1\over 2} \left [\te{s}{_i^j}s_{jk}+\sigma^2\left (u_iu_k +{g_{ik}\over 2}\right )
\right ]
\label{U1}
\ee 
in the Einstein-Cartan equations
\be
G_{ik}=P_{ik}-\left (\lambda+\frac P 2\right ) g_{ik}=\kappa {T}_{ik}+\kappa \varepsilon Z_{ik}+\kappa^2 U_{ik}.
\label{EC}
\ee
Here we use $P=\te{P}{_i^i}$ and the cosmological constant $\lambda$. An additional part by torsion comes from the Belinfante-Rosenfeld equation \cite{BEL40,Ros40} which relates the dynamical metric ${\cal T}$ and the canonical energy-momentum $T$ tensor by 
\be
{\cal T}_{ik}&=&T_{ik}+\varepsilon Z_{ik}
\nonumber\\
Z_{ik}&=&
-\frac 1 2 \nabla_l (\te{s}{_k^l} u_i+\te{s}{_i^l}u_k+\te{s}{_k_i}u^l).
\label{BR}
\ee
This difference we indicate by a constant $\epsilon=0$ or $\epsilon=1$ dependent whether we use the metric variational principle or the torsion tensor as variation variable. The metric and torsion are dependent on each other even outside matter which consistency results into the antisymmetric spin tensor in terms of the Weyssenhoff spin liquid parameter $2 \sigma^2=\te{s}{^m_l}\te{s}{_m^l}$ with the only nonzero components in spherical coordinates \cite{Mo21}
\be
s_{2,3}(t,r,\theta, \phi)=-s_{3,2}=\sigma r^2 \sin\theta 
.
\ee
Besides the Schwarzschild solution with zero torsion it has been found that there exists exactly a second solution of these equations (\ref{EC}) outside matter withe the metric \cite{Mo21}
\ba
ds^2&=
\hat r^2 d\hat t^2\!-\!{3 \over C\!+\! \lambda \hat r^2} d \hat r^2
\!-\!\hat r^2(d\theta^2\!+\!\sin^2\theta d\hat \phi^2)
\label{metricd}
\end{align}
where we abbreviate 
\be
C={3 (1+2 \varepsilon)\over 2(2+3\varepsilon)}
=\left \{\begin{array}{ll}
3/4 & \epsilon=0\cr
{9/10} & \epsilon=1
\end{array}\right . .
\label{Con}
\ee
A further transformation 
\ba
&\hat r^2=\cos^2 \tilde t \left [c(\tilde r)^2\tan^2\tilde t -1\right ],\,
\coth\hat t=c(\tilde r) \tan \tilde t
\end{align}
with 
\ba
c(\tilde r)={\rm tan} \left (c+ \sqrt{C\over 3} \,{\rm ln}\, \tilde r\right ),
\quad \tilde r=\sqrt{|\lambda|\over 3} r,
\quad \tilde t=\sqrt{|\lambda|\over 3} t
\label{itrafo3}
\end{align}
translates the metric (\ref{metricd}) into a Friedman-Lama\^itre-Robertson-Walker metric 
\be
ds^2=d t^2-a(r, t)\left (d r^2+r^2 d\Omega^2\right ).
\label{ds}
\ee
The expansion or scale parameter becomes now space and time dependent
\be
a(r, \tilde t)=R^2(r,t)={C \sin^2{\sqrt{|\lambda|\over 3}t}\over |\lambda| r^2\cos^2{\left (2 c+ \sqrt{C\over 3} {\rm ln} \,\sqrt{|\lambda|\over 3}r\right )}}
\label{Ryx1}
\ee
with an arbitrary constant $c$. It provides a time-like universe for ${\tilde r}^2<\coth^2 \tilde t-1$ and a space-like otherwise.

\section{Time dependence of Hubble parameter}

\subsection{Local simplistic picture}
\begin{figure}
\centerline{\includegraphics[width=8.5cm]{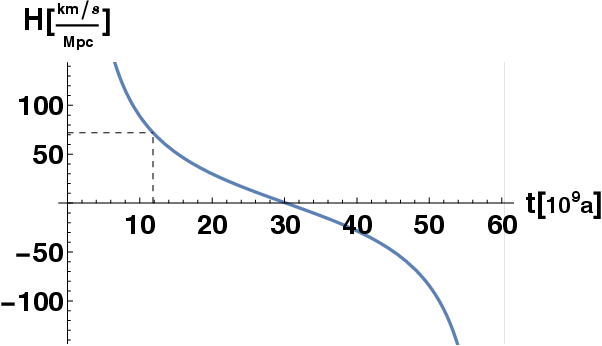}}
\caption{\label{hubble} Simplistic (locally fixed) time dependence of Hubble constant (\ref{Ht}) leading to the wrong time dependence.
Zero time is set at vanishing Hubble constant and the present is indicated by dashed lines.}  
\end{figure}

From the seemingly factorization of time and space dependence in the expansion parameter (\ref{Ryx1}) one might be tempted to calculate  locally the Hubble constant $H$ and the delay parameter $q$ directly via (\ref{hq})
as
\be
H(t)&=&{\dot R\over R}=\sqrt{|\Lambda|\over 3}\cot\sqrt{|\Lambda|\over 3} t 
\newline\\
q(t)&=&-{R {\ddot R}\over {\dot R}^2}=\tan^2\sqrt{|\Lambda|\over 3} t ={\lambda\over 3 H^2(t)}
\label{Ht}
\ee
which divides out the spatial dependence which is too simplistic and will lead to a wrong behaviour. This result is plotted in figure~\ref{hubble}. We see that we have only a time-decreasing Hubble parameter with $q_0>0$ in contrast to the data above. The reason is that the light coming from the past is running on $r=r_0+c (t-t_0)$ and we cannot just simply divide out the spatial dependence. This we will consider more closely in the next chapter. 

First let us see to what values the local assumption (\ref{Ht}) will lead which might explain a discrepancy to earlier data. We assume that the starting time $t_0$ is fixed at $H(t_0)=0$. Then the Hubble constant at present time $H_0=H(t_p)$ and $q_0$ is related with the cosmological constant as 
\be
\lambda=3 q_0 H_0^2\sim H_0^2
\ee 
which is in agreement of high-redshift rotation curves and MOND calculations \cite{milgrom17}. Within this simplistic picture we assume the older value $q_0\approx 1/2$ and see from (\ref{Ht})
that the initial time is $t_0=0$ for $H=\infty$, the present and final time where the Hubble constant vanishes read
\be
t_p
&=&{1\over \sqrt{q_0}H_0}
(\pi-{\rm arccot}{\sqrt{q_0}})-t_0\approx 11.8\times 10^9 {\rm a}
\nonumber\\
t_\infty&=&{\pi\over \sqrt{q_0}H_0}-t_0\approx 30.2\times 10^9 {\rm a},
\ee
respectively, with $1/H_0\approx 13.6\times 10^9 a$.

The relative time change of the Hubble constant plotted in figure~\ref{hubble} in this simplistic picture would be at present
${\dot H/ H}(t_0)\approx -2\times 10^{-10}{\rm a}^{-1}$.
All these times agree astonishingly well with the Big Bang scenario though only locally measured and with the wrong time dependence. We will see that these values changes only slightly if we consider correctly the time the light travels from far distances which will lead to the right time dependence of accelerating Hubble parameter. The wrong time dependence of the Hubble parameter here in local approximation has followed only from the structure of the metric and the cosmological constant and is independent of the factor $C$ describing the spatial dependence of the expansion parameter. This problem in late-time cosmology are typical for models loke Lambda cold dark matter ($\Lambda$CDM).  Extensions lead to models with an extra parameter \cite{Mae17}, or e.g. investigation of unparticle cosmology \cite{AP21}.

\subsection{Hubble parameter from distant light}

\begin{figure}
\centerline{\includegraphics[width=8.5cm]{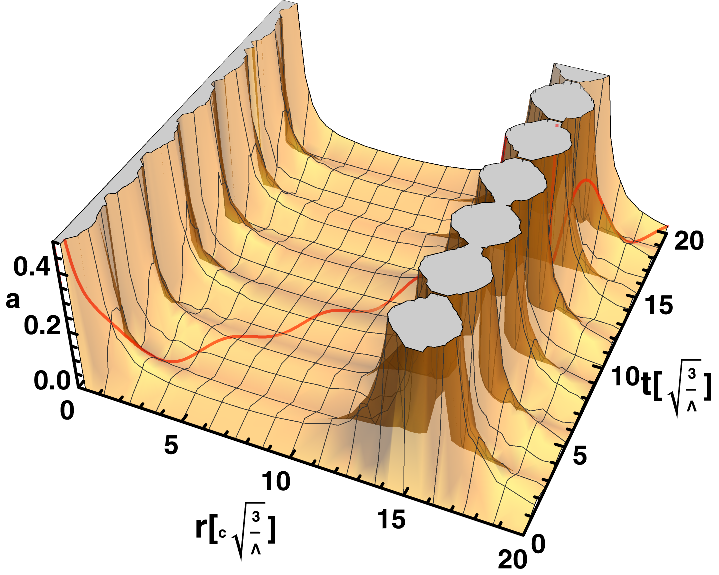}
}
\caption{\label{a_rt} The time and space dependence of the expansion parameter (\ref{Ryx1}) with the light path $r=r_0+c (t-t_0)$ (red).}  
\end{figure}

The spatial and time dependence of the expansion parameter (\ref{Ryx1}) allows to consider the time the light is traveling from far distances. In figure~\ref{a_rt} we plot the square of this expansion parameter as a function of time and space. Contrary to the simplistic picture before we measure light from distant objects. This means we cannot consider space and time independently as before. Instead we have to consider Hubble parameter at the light path $r=r_0+c (t-t_0)$ indicated by the red line in figure~\ref{a_rt}. One sees the oscillating behaviour with respect to space and time. The spatial variation shows interestingly one additional maxim at large distances before it falls off rapidly. 

Working in dimensionless values
\be
H=\sqrt{\lambda \over 3} h,\quad \bar r =\sqrt{\lambda \over 3 c^2} r,\quad \tau=\sqrt{\lambda \over 3} t
\ee
with speed of light $c$ we obtain with time derivatives along the light path the time dependent Hubble and deceleration parameter
\be
h(\tau)&=&\cot \tau+{\sqrt{C \over 3}\tan{\sqrt{C\over 3}\ln({\bar r_0}+\tau)}-\sqrt{3}
\over {\bar r_0}+\tau}
\nonumber\\
q(\tau)&=&-1+{1
\over
h ({\bar r_0}+t)}
+{({\bar r_0}+t) \csc^2 t-\cot t
\over
 h^2 ({\bar r_0}+t)}
-
{C
\sec^2\left(
\sqrt{C \over 3}\log ({\bar r_0}+t)
\right)
\over
3 h^2 ({\bar r_0}+t)^2}.
\label{hq1}
\ee

\begin{figure}
\centerline{\includegraphics[width=8.5cm]{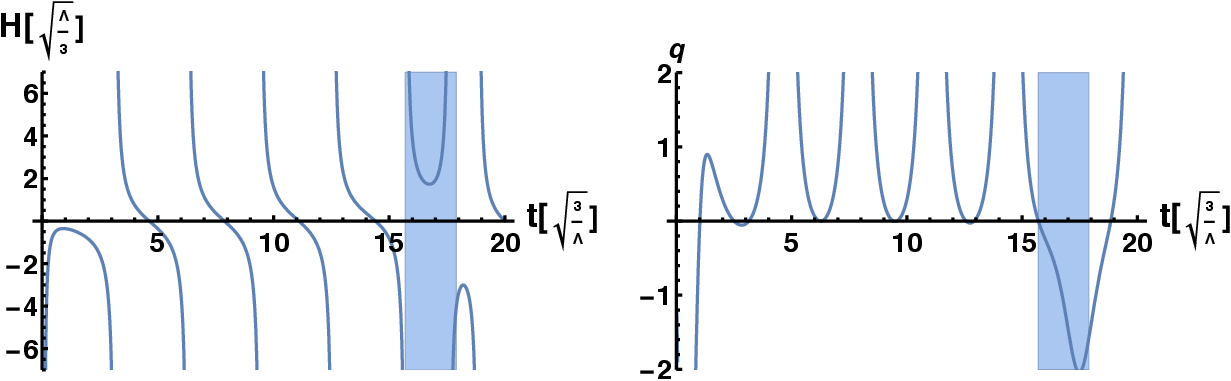}}
\centerline{\includegraphics[width=8.5cm]{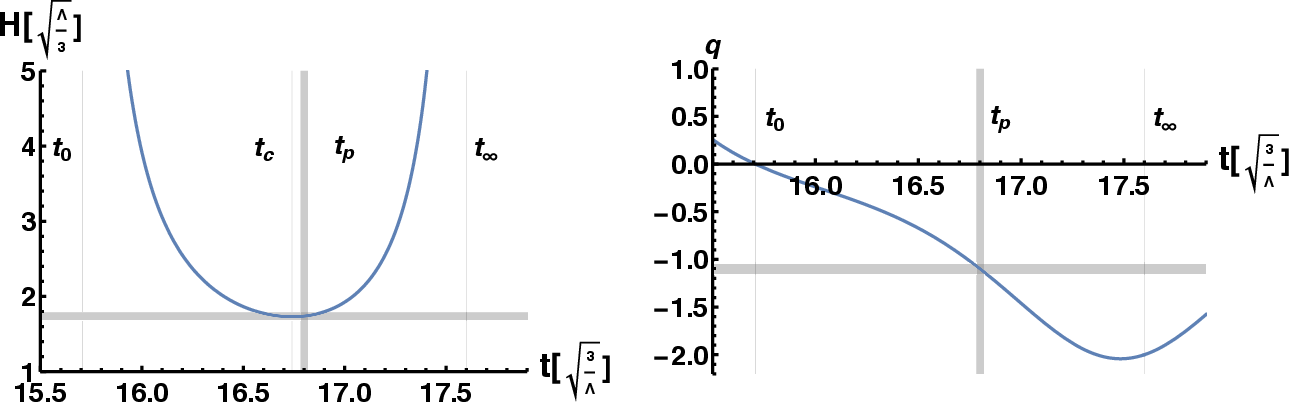}}
\caption{\label{hq_c} The dimensionless Hubble parameter (left) and deceleration parameter (right) as function of dimensionless time assuming a present position of $r_0=0$. Below a zoom of the only possible window with $\dot h>0$ indicated by shaded area. The present values at $t=t_p$ according to (\ref{values}) are indicated by thick grid-lines.}  
\end{figure}

\begin{figure}
\centerline{\includegraphics[width=8.5cm]{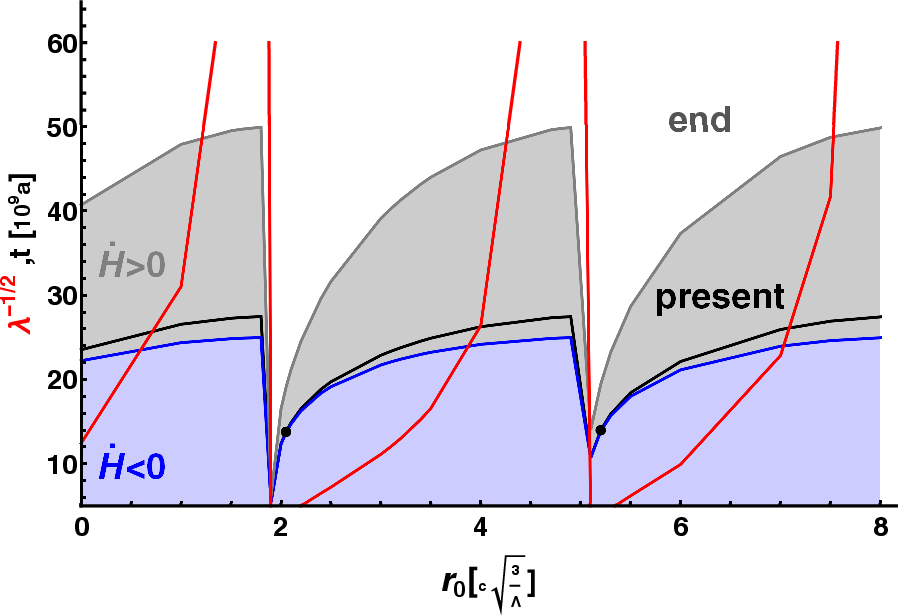}}
\caption{\label{age_t} The age of present universe together with the age where the Hubble parameter changes from falling into increasing value vs. the dimensionless parameter of present location $r_0$. The data at presence are middle (black) lines and the corresponding cosmological constant are red lines. The best agreement with experimental data are indicated by dots outlines in figure~\ref{Hz}.}  
\end{figure}

In figure~\ref{hq_c} we see that the periodic oscillation along the light cone reveals only one certain window where the Hubble parameter can increase with time  $\dot H>0$ as observed. Choosing this interval as indicated by the shaded area we can determine the initial and final time of this universe window as the present cosmos by $H(t_0)=H(t_\infty)=\infty$, the present time by $H(t_p)$
and the time where the Hubble parameter changes from falling to increasing time by $\dot H(t_c)=0$. We determine the present time by reproducing the deceleration parameter (\ref{hq1}) according to the value (\ref{values}).
For any parameter $r_0$ we can now determine these times together with the cosmological constant plotted in figure~\ref{age_t}.
  
We see that the unknown parameter $\bar r_0$ as the starting place in figure~\ref{a_rt} determines all three values of initial time, ending time as well as the cosmological constant due to the known present data (\ref{values}). Larger $\bar r_0$ implies larger times accompanied by larger cosmological constant. In turn if we know the cosmological constant by other measurements we know $\bar r_0$ and the times are fixed. Please note that we have set the timescale to initially $t_0=0$ such that only the differences in times matter. The oscillating behaviour as big bounce instead of big bang has been reported in \cite{POP10,Pop12} due to torsion. Such turning point in the Hubble parameter was obtained by considering unstable de Sitter state \cite{CMY19}. Dependent on the parameter $r_0$ we can now find different ages and ending of the universe.

\begin{figure}
\centerline{\includegraphics[width=4.2cm]{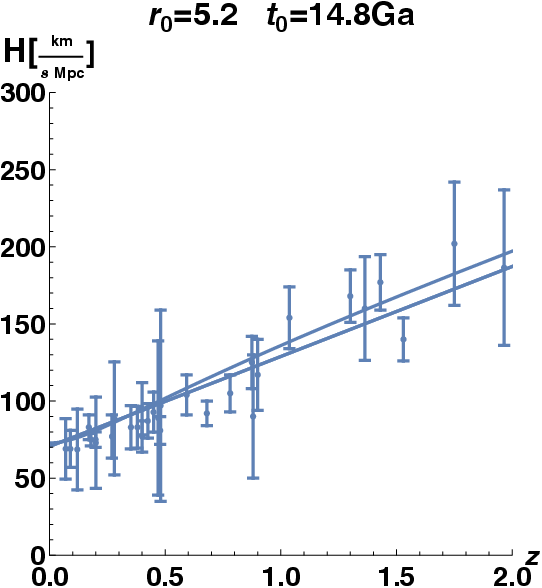}\includegraphics[width=4.2cm]{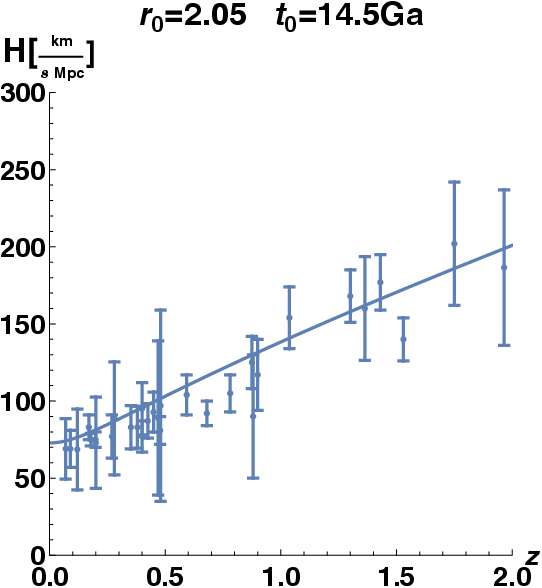}}
\caption{\label{Hz} The Hubble parameter versus redshift together with the experimental data compilation of \cite{RDR18} for the best situation of figure~\ref{age_t}.}  
\end{figure}

For a given $r_0$ parameter we have the time dependence of Hubble parameter and deceleration parameter in (\ref{hq1}) fixed according to (\ref{values}). We can now compare with the experimental values which are given in terms of the red shift $z=-1+R_0/R$. One has for the time dependence
\be
\dot z=-{R_0\over R}{\dot R\over R}=(-1+z) H
\ee
and therefore with $z_0=0$ at present time $t_0$
\be
z(t)=-1+\exp{\left [-\int\limits_{t_0}^t d\bar r H(\bar t)\right ]}.
\ee 
Using the time as parameter we can plot in figure~\ref{Hz} the Hubble parameter versus redshift keeping the present Hubble constant at the value of (\ref{values}). The best choices of $r_0$ are indicated by dots in figure~\ref{age_t} according to the experimental data in figure~\ref{Hz}. For further comparison of the data with present models see \cite{AP21}. The resulting time where decelerating Hubble parameter changes into accelerating $t_c$, the present age $t_p$ and the end of universe $t_\infty$ read then
\be
r_0&=&2.05\, (4.1 Gpc):\nonumber\\
t_c&=&14.0413 Ga,\quad 
t_0=14.1518 Ga, \quad 
t_\infty=19.5727 Ga,\quad 
\lambda^{-1/2}=3.78811 Ga\nonumber\\
 r_0&=&5.2\,  (10.7 Gpc):\nonumber\\
t_c&=&14.2399 Ga,\quad  
t_0=14.3581 Ga,\quad  
t_\infty=19.9614 Ga,\quad 
 \lambda^{-1/2}=3.87425 Ga.
\label{struc}
\ee
In figure~\ref{age_t} one sees that almost similar realizations of experimental values are possible for periodically appearing $r_0$ where we have listed only the first two ones.

\section{Conclusion}

A time dependence of the Hubble and deceleration parameter is found from the exact solution of Einstein-Cartan equations. The latter provides a spatial and time dependent expansion or scale parameter. It is shown that the evolution of the universe is starting with an decreasing Hubble parameter switching to an increasing one within a certain evolution window among possible cosmoses. The seemingly dependence of the certain times and Hubble behaviour on the position parameter $r_0$ is indicating a violation of equivalence principle. It appears here as an artificial unknown parameter for large scale structures determined by the cosmological constant. Since the Einstein-Cartan equations complete the equivalence principle \cite{Hey75} or more recently \cite{Pra22} and since we have used an exact solution of the latter we can conclude that locally there exist a transformation to a frame where the gravitational force vanishes though the large scale time and space structure of the expansion parameter looks nonholonomic.

As a second hint we should note that the torsion can mime dark matter. This can be seen as follows. We rewrite (\ref{metricd}) into the Schwarzschild form:
\be
ds^2=\left (1-\frac a r\right ) d t^2-{1\over 1-\frac b r} dr^2-r^2 d\Omega^2
\ee
with
\be
a=r-\frac {|\Lambda|}{ 3} r^3,\qquad b=\left (1-\frac C 3 \right ) r-\frac {|\Lambda|} 3 r^3.
\ee 
If we compare this result of the new metric with
the standard Schwarzschild solution with zero torsion and the extension to include the cosmological constant known as Kottler solution \cite{CXZ87}
\be
a^K=2 M-\frac {|\Lambda|}{ 3} r^3,\qquad b^K=2 M -\frac {|\Lambda|}{ 3} r^3
\ee 
we can conclude that the new metric resulting from torsion induces a mass like term
\be
M^{\rm torr}=\frac 1 2 \left (1-\frac C 3 \right ) r
\ee
which increases with larger distances. This can probably mime an additional gravitational mass \cite{Pop11} modifying the outer rotation of large galaxies \cite{mor21}. Recent investigations for torsion leading to dark energy can be found in \cite{ben22}. 

The found periodic spatial dependence $r_0$ in (\ref{struc}) agrees well with the recently observed large ring structures \cite{LCW22, Ku23}.



\begin{thebibliography}{10}

\bibitem{Riess22}
A.~G. Riess, W.~Yuan, L.~M. Macri, D.~Scolnic, D.~Brout, S.~Casertano, D.~O.
  Jones, Y.~Murakami, G.~S. Anand, L.~Breuval, T.~G. Brink, A.~V. Filippenko,
  S.~Hoffmann, S.~W. Jha, W.~D. Kenworthy, J.~Mackenty, B.~E. Stahl and
  W.~Zheng, {\em The Astrophysical Journal Letters} {\bf 934},  ~L7 (jul 2022).

\bibitem{RL19}
R.~G and E.~Lusso, {\em Nat. Astron.} {\bf 3},   172  (2019), arXiv:1811.02590.

\bibitem{Li21}
X.~Li, R.~E. Keeley, A.~Shafieloo, X.~Zheng, S.~Cao, M.~Biesiada and Z.-H. Zhu,
  {\em Monthly Notices of the Royal Astronomical Society} {\bf 507}, 919 (07
  2021).

\bibitem{CJSBCDS22}
E.~\'O~Colg\'ain, M.~M. Sheikh-Jabbari, R.~Solomon, G.~Bargiacchi,
  S.~Capozziello, M.~G. Dainotti and D.~Stojkovic, {\em Phys. Rev. D} {\bf
  106},   L041301 (Aug 2022).

\bibitem{MCCPS23}
M.~Malekjani, R.~M. Conville, E.~. Colg\'ain, S.~Pourojaghi and M.~M.
  Sheikh-Jabbari  (2023), {{
  arXiv:2301.12725 [astro-ph]}}.

\bibitem{Riess19}
A.~G. Riess, S.~Casertano, W.~Yuan, L.~M. Macri and D.~Scolnic, {\em The
  Astrophysical Journal} {\bf 876},  ~85 (may 2019).

\bibitem{CM20}
D.~Camarena and V.~Marra, {\em Phys. Rev. Research} {\bf 2},   013028 (Jan
  2020).

\bibitem{CMY19}
E.~{O Colg\'ain}, M.~H. {van Putten} and H.~Yavartanoo, {\em Physics Letters B}
  {\bf 793}, 126  (2019).

\bibitem{MZJ20}
W.-M. Dai, Y.-Z. Ma and H.-J. He, {\em Phys. Rev. D} {\bf 102},   121302 (Dec
  2020).

\bibitem{CJ21}
E.~. Colg\'ain and M.~M. Sheikh-Jabbari, {\em Classical and Quantum Gravity}
  {\bf 38},   177001 (aug 2021).

\bibitem{Mo21}
K.~Morawetz, {\em Classical and Quantum Gravity} {\bf 38},   205003 (sep 2021).

\bibitem{Mo21e}
errata, {\em Class. Quantum Grav.} {\bf 40},   029501  (2022).

\bibitem{BEL40}
F.~Belinfante, {\em Physica} {\bf 7}, 449   (1940).

\bibitem{Ros40}
L.~Rosenfeld, {\em Sur le tenseur d'impulsion-{\'e}nergie} (Palais des
  Acad{\'e}mies (impr. de G. Thone), Bruxelles, 1940).

\bibitem{milgrom17}
M.~Milgrom, High-redshift rotation curves and mond  (2017).

\bibitem{Mae17}
A.~Maeder, {\em The Astrophysical Journal} {\bf 834},   194 (jan 2017).

\bibitem{AP21}
M.~A. Abchouyeh and M.~H. P.~M. van Putten, {\em Phys. Rev. D} {\bf 104},
  083511 (Oct 2021).

\bibitem{POP10}
N.~J. Popławski, {\em Physics Letters B} {\bf 694}, 181   (2010).

\bibitem{Pop12}
N.~Pop\l{}awski, {\em Phys. Rev. D} {\bf 85},   107502 (May 2012).

\bibitem{RDR18}
J.~Ryan, S.~Doshi and B.~Ratra, {\em Mon. Not. R. Astron. Soc.} {\bf 480},
  759  (2018).

\bibitem{Hey75}
P.~Von Der~Heyde, {\em Lett. Nuovo Cimento} {\bf 14},   250  (1975).

\bibitem{Pra22}
G.~Pradisi and A.~Salvio, {\em Eur. Phys. J. C} {\bf 82},   840  (2022).

\bibitem{CXZ87}
X.~Chongming, W.~Xuejun and H.~Zhun, {\em Gen. Relat. Gravit.} {\bf 19},   1203
   (1987).

\bibitem{Pop11}
N.~J. Pop\l{}awski, {\em Phys. Rev. D} {\bf 83},   084033 (Apr 2011).

\bibitem{mor21}
A.~L. Gonz\'alez-Mor\'an, R.~Ch\'avez, E.~Terlevich, R.~Terlevich,
  D.~Fern\'andez-Arenas, F.~Bresolin, M.~Plionis, J.~Melnick, S.~Basilakos and
  E.~Telles, {\em Monthly Notices of the Royal Astronomical Society} {\bf 505},
  1441 (05 2021).

\bibitem{ben22}
D.~Benisty, E.~Guendelman and H.~Stoecker, {\em The European Physical Journal}
  {\bf 82},   264  (2022).

\bibitem{LCW22}
A.~M. Lopez, R.~G. Clowes and G.~M. Williger, {\em Monthly Notices of the Royal
  Astronomical Society} {\bf 516}, 1557 (08 2022).

\bibitem{Ku23}
P.~K. Aluri, P.~Cea, P.~Chingangbam, M.-C. Chu, R.~G. Clowes, D.~Hutsemekers,
  J.~P. Kochappan, A.~M. Lopez, L.~Liu, N.~C.~M. Martens, C.~J. A.~P. Martins,
  K.~Migkas, E. \'O. Colg\'ain, P.~Pranav, L.~Shamir, A.~K. Singal, M.~M.
  Sheikh-Jabbari, J.~Wagner, S.-J. Wang, D.~L. Wiltshire, S.~Yeung, L.~Yin and
  W.~Zhao, {\em Classical and Quantum Gravity} {\bf 40},   094001 (apr 2023).

 \end{thebibliography}

\end{document}